\documentclass{article}
\usepackage[utf8]{inputenc}
\usepackage[english]{babel}
\usepackage{amsmath}
\usepackage{amsfonts}
\usepackage{amssymb}
\usepackage{graphicx}
\usepackage{subcaption}
\usepackage{authblk}
\usepackage{hyperref}
\providecommand{\keywords}[1]
{
  \small	
  \textbf{\textit{Keywords }} #1
}

\title{COVID-Pro in Italy: \\ A dashboard for a province-based analysis}
\author[1]{Luisa Ferrari}
\author[1]{Giuseppe Gerardi}
\author[1]{Giancarlo Manzi}
\author[2]{Alessandra Micheletti}
\author[1]{Federica Nicolussi}
\author[1]{Silvia Salini}

\affil[1]{Department of Economics, Management and Quantitative Methods, University of Milan, Italy}
\affil[2]{Department of Environmental Science and Policy,  University of Milan, Italy}

\date{Last updated: April 27th, 2020}

\begin{document}

\maketitle

\begin{abstract} This paper presents an dashboard developed to analyse the outbreak of the Covid-19 infection in Italy considering daily NUTS-3 data on positive cases provided by the Italian Ministry of Health and on deaths derived from Italian regional authorities' official press conferences. Descriptive time series plots are provided together with a map describing the spatial distribution of province cumulative cases and rates. A section on a proposed time-dependent adjusted SIRD model for NUTS-3 regions is also provided in the dashboard.
\end{abstract}

\keywords{Covid-19, Italy, NUTS-3 regions, SIRD model}

\section{Introduction}
This dashboard (\url{http://demm.ceeds.unimi.it/covid/}) is focused on the analysis of the SARS-Cov2 outbreak in Italy at a \textit{provincial} (i.e. EU NUTS-3) level. It has been developed using \texttt{Shiny}, an easy web application in \texttt{R} from RStudio Inc, (\url{http://www.rstudio.com/shiny/}).

Our choice to focus on Italian provinces, rather than on Italian regions (i.e. at EU NUT-2 level), has been dictated by the fact that the outbreak of SARS-Cov2 in Italy has been so far not homogeneously spread within regions, with many differences from province to province.

According to the characteristics of the virus spreading, it seems difficult to think of a homogeneous propagation behavior at the regional level. Even the timing of the initial stages of the infection and its dynamics seem to have been very different even among contiguous provinces and inside the provinces themselves, as clusters of Covid-19 contagion have been often located in very restricted areas. 

Antigen testing had been initially conducted depending on the choices of the local health authorities, and hospital admissions in the early stages of the emergency depended on the management and absorption capacity of the local health units. 

Only data on total cases are currently available at a provincial level. We integrated these data deriving the provincial Covid-19 deaths from press reports downloaded from regional authorities.

\subsection{Provincial deaths data collection}
\noindent
The official data repository of the Italian Ministry of Health and the Civil Protection Agency does not provide Covid-19 data on the daily number of deaths at a provincial level. However,by scraping the daily press conferences and Covid-19 bulletins provided by 13 regions out of 20, we were able to obtain provincial data on a daily basis regarding the number of deaths due to the Covid-19 infection for a vast majority of the Italian provinces. The Aosta province and the provinces of Bolzano/Bozen and Trento are included in the Italian Ministry of Health regional data as they are either coincident with the region (Valle d'Aosta) or are autonomous provinces (Bolzano/Bozen and Trento) and considered as regions. In Table \ref{tab1} for each Italian region we provide the main source where we scraped these data. 

Regions for which we were not able to obtain provincial death data from official bulletins or press conferences were Liguria, Lombardia, Veneto, Friuli-Venezia-Giulia and Campania. However, for Imperia in the Liguria region and for Lodi and Cremona provinces in the Lombardia region we were able to obtain data on Covid-19 deaths from local newspapers. Table \ref{tab1} contains all the sources we scraped for obtaining provincial Covid-19 death data.

\begin{table}[]
 \centering
    \begin{tabular}{|l|l|}
          \hline
          Region & Main source \\
          \hline
          Valle d'Aosta & https://github.com/pcm-dpc/COVID-19  \\
          Piemonte & https://www.regione.piemonte.it/web/pinforma\\
          Lombardia & https://www.ilcittadinodilodi.it and \\
          & https://laprovinciacr.it  (data only for Lodi \\
          & and Cremona provinces)\\
          Veneto & no available data\\
          Friuli-Venezia-Giulia & no available data\\
          Trentino-Alto-Adige & https://github.com/pcm-dpc/COVID-19 \\
          Emilia-Romagna & https://www.regione.emilia-romagna.it/notizie/2020 \\
          Liguria & http://sanremonews.it (data only for Imperia province) \\
          Toscana & https://www.toscana-notizie.it/coronavirus \\
          Marche & http://www.regione.marche.it/News-ed-Eventi \\
          Umbria & http://www.regione.umbria.it\\
          Lazio & https://www.facebook.com/SaluteLazio \\
          Abruzzo & https://www.regione.abruzzo.it/notizie\_stampa\\
          Molise & http://www3.regione.molise.it\\
          Campania & no available data\\
          Puglia & http://www.regione.puglia.it \\
          Basilicata & https://www.regione.basilicata.it \\
          Calabria & https://portale.regione.calabria.it \\
          Sicilia & http://pti.regione.sicilia.it \\
          Sardegna & https://www.regione.sardegna.it \\
     \hline
     \end{tabular}
    \caption{Main data sources for provincial covid-19 deaths}
     \label{tab1}
 \end{table}
 
\subsection{List of provinces with Covid-19 provincial deaths}
 The following is the list of Italian provinces for which we obtained deaths data.
 
 Agrigento, Alessandria, Ancona, Aosta, Arezzo, Ascoli Piceno, Asti, Bari, Barletta-Andria-Trani, Biella, Bologna, Bolzano, Brindisi, Cagliari, Caltanissetta, Campobasso, Catania, Catanzaro, Chieti, Cosenza, Cremona, Crotone, Cuneo, Enna, Fermo, Ferrara, Firenze, Foggia, Forli-Cesena, Frosinone, Grosseto, Imperia, Isernia, L'Aquila, Latina, Lecce, Livorno, Lodi, Lucca, Macerata, Massa Carrara, Matera, Messina, Modena, Novara, Nuoro, Oristano, Palermo, Parma, Perugia, Pesaro e Urbino, Pescara, Piacenza, Pisa, Pistoia, Potenza, Prato, Ragusa, Ravenna, Reggio di Calabria, Reggio nell'Emilia, Rieti, Rimini, Roma, Sassari, Siena, Siracusa, Sud Sardegna, Taranto, Teramo, Terni, Torino, Trapani, Trento, Verbano-Cusio-Ossola, Vercelli, Vibo Valentia, Viterbo

\section{Province maps}
\begin{figure}[ht!]
    \centering
    \begin{subfigure}[c]{0.4\textwidth}
        \includegraphics[width=\textwidth]{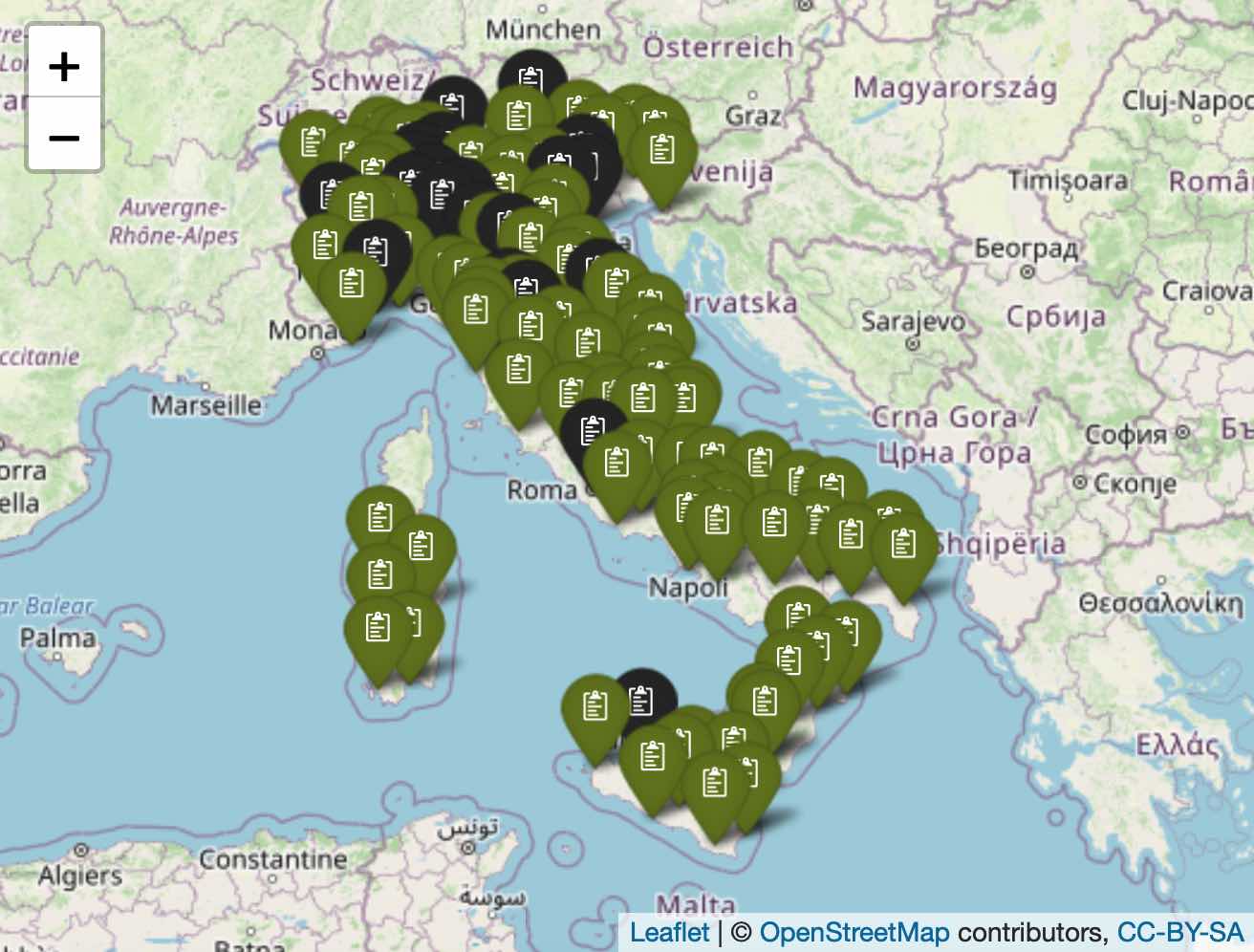}
        \caption{Cumulative rates late February in Italy}
    \end{subfigure}
    \begin{subfigure}[c]{0.4\textwidth}
        \includegraphics[width=\textwidth]{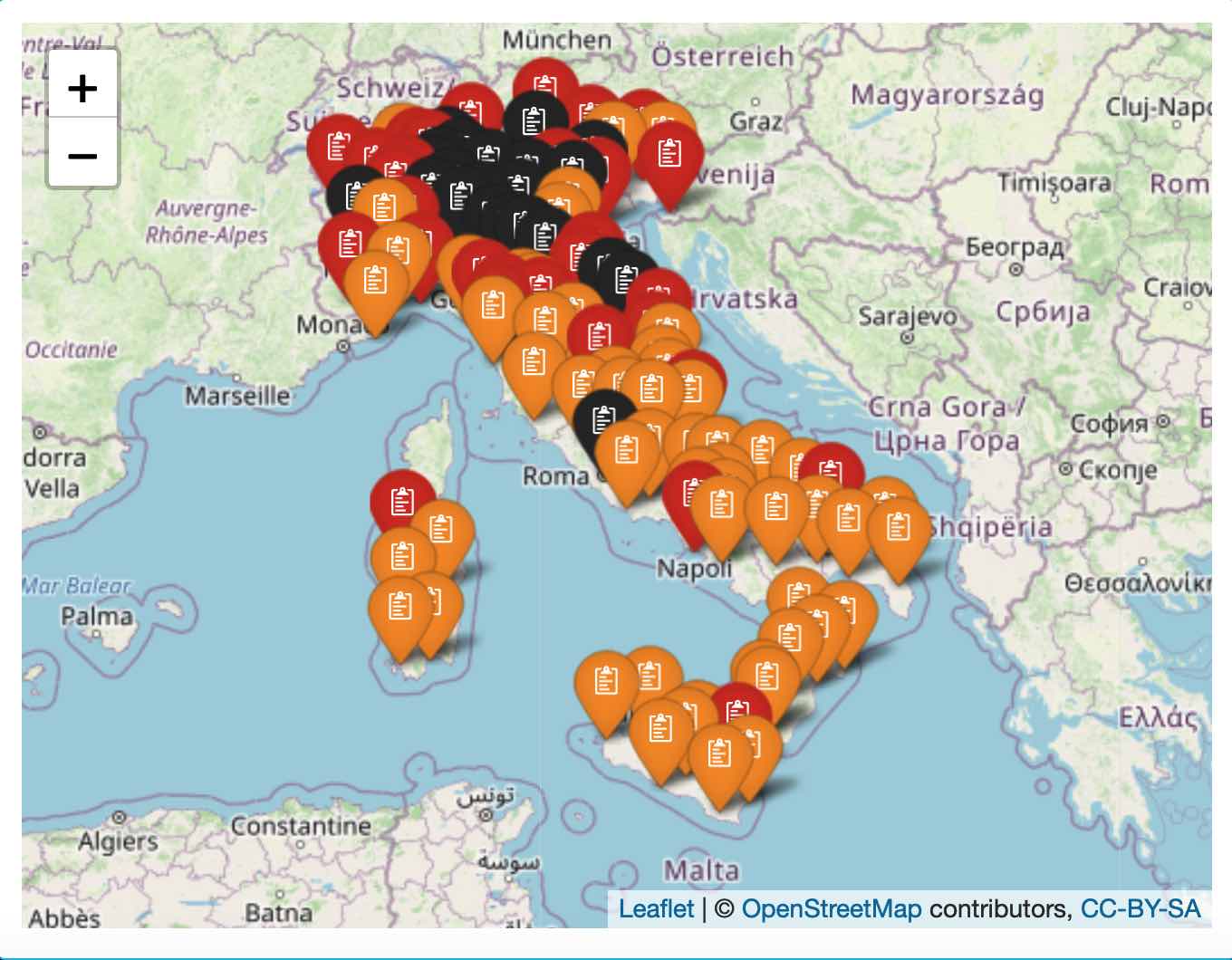}
        \caption{Cumulative rates late April in Italy}
    \end{subfigure}
    \caption{Maps for provincial cumulative rates in Italy - Late February and April, 2020}
    \label{fig:fig1}
\end{figure}

From the dashboard, it is possible to see a map of Italy with provinces highlighted according to their cumulative Covid-19 cases and rates per 100,000 inhabitants (Figure \ref{fig:fig1}), the latter being computed as\footnote{Sources: Italian Ministry of Health for cases, Italian National Statistical Institute (ISTAT) for the population.}:

\begin{align*}
\frac{Cases}{Population} * 100000
\end{align*}

\noindent From the main menu it is possible to switch between cumulative cases and cumulative rates and choose the reference period.
Figure \ref{fig:fig1} shows the situation in late February and late April for cumulative rates.

\begin{figure}[ht!]
    \centering
    \begin{subfigure}[c]{0.4\textwidth}
        \includegraphics[width=\textwidth]{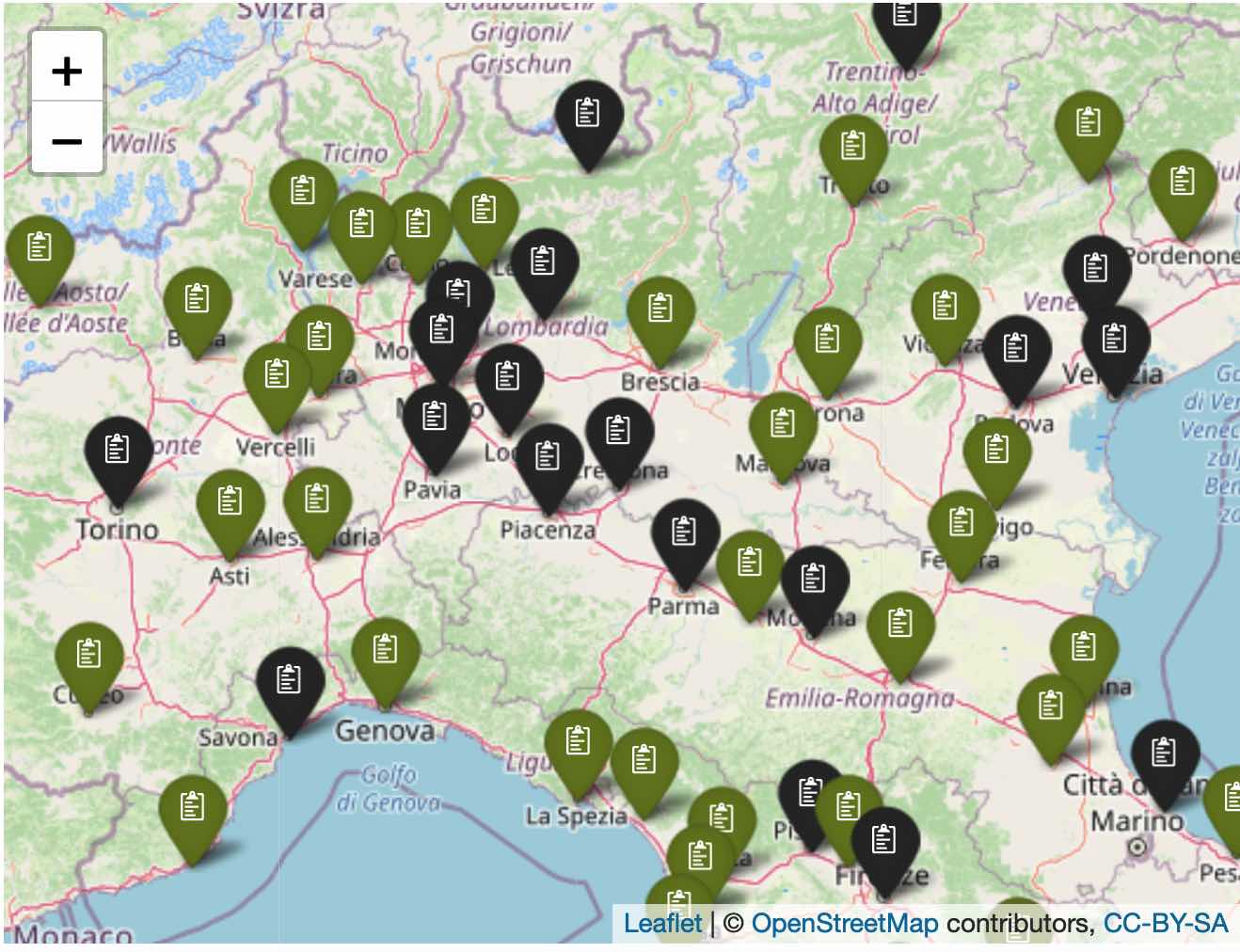}
        \caption{Cumulative rates in late February in Italy}
    \end{subfigure}
    \begin{subfigure}[c]{0.4\textwidth}
        \includegraphics[width=\textwidth]{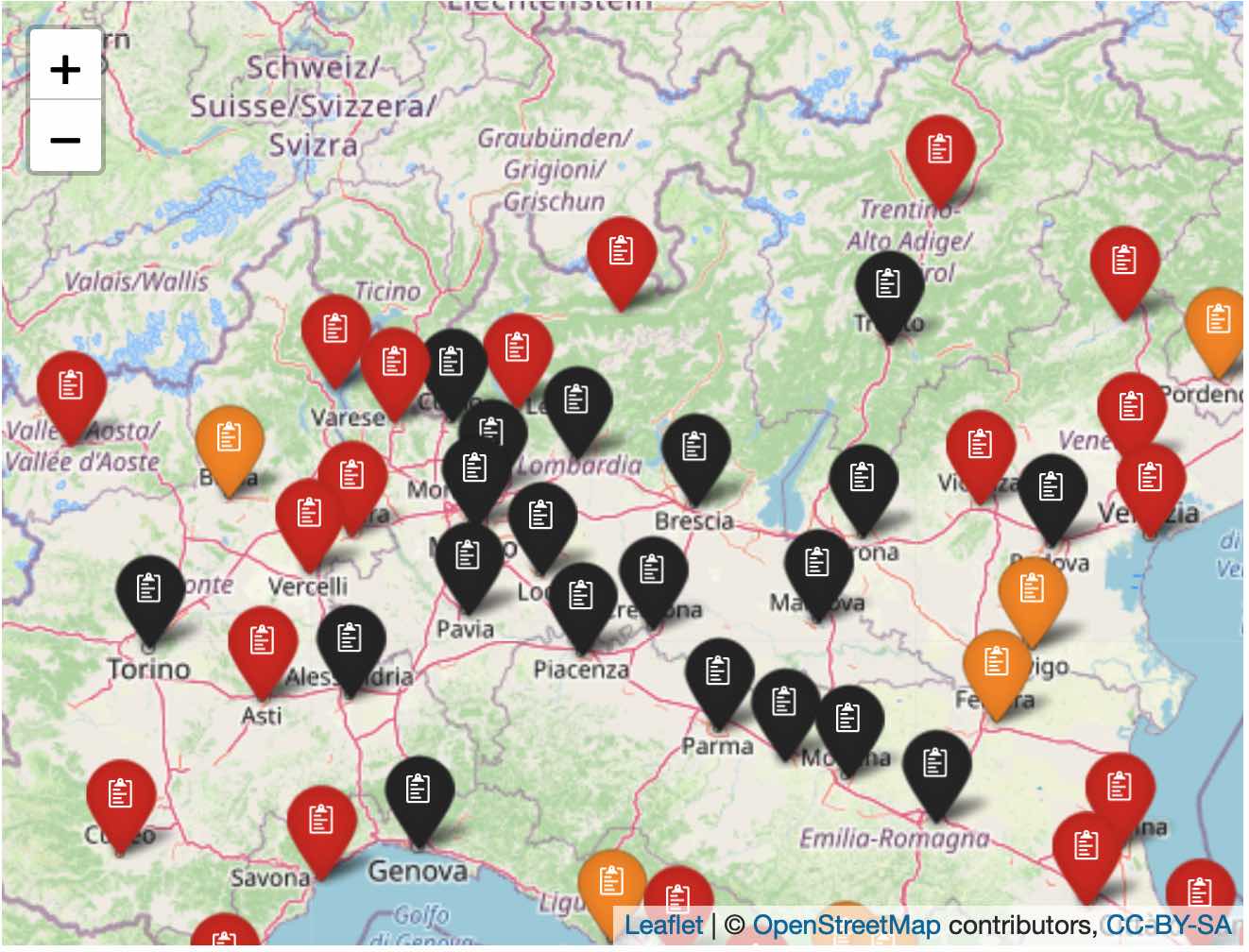}
        \caption{Cumulative rates in late April in Italy}
    \end{subfigure}
    \caption{Maps for the initial 'red zone'}
    \label{fig:fig2}
\end{figure}

\noindent By zooming individual areas, cumulative rates or cases relating to provinces can be seen. Figure \ref{fig:fig2} shows the most affected areas in Northern Italy in late February and late April, the so-called 'initial red-zone'. The colors represent the intensity of cumulative cases or rates, from green (0\%), orange (lower than 50\%), red (between 50\% and 80\%), black (higher than 80\%)

\section{Time series}

\subsection{Deaths rate by province} 

Death rates are the ratio between the cumulative death cases and the total population for the selected provinces multiplied by 100,000. To build this time series we used data collected by scraping the daily press conferences and Covid-19 bulletins provided by regions, as the official data repository of the Italian Ministry of Health and the Civil Protection Agency does not provide Covid-19 data on the daily number of deaths at a provincial level, but only on a regional level.

\begin{figure}[ht]
    \centering
    \includegraphics[scale=0.4]{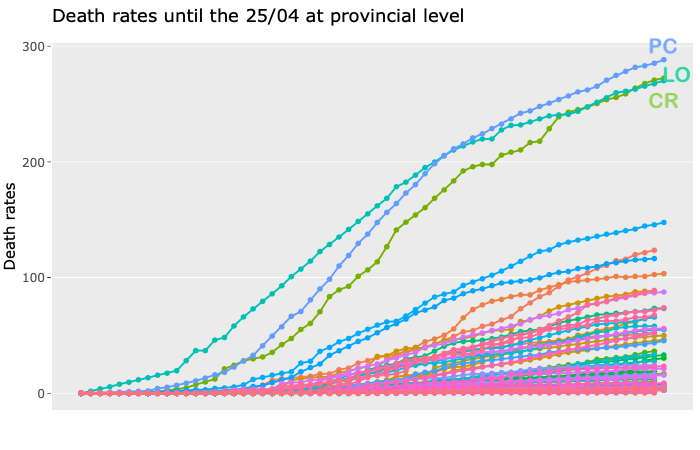}
    \caption{Cumulative death rates}
    \label{fig:fig3}
\end{figure}

Figure \ref{fig:fig3} shows the cumulative death rate by province. According to \cite{Signorelli} the provinces with the largest value are Piacenza, Lodi and Cremona, it is possible to see the province and the value by clicking on the points. 

Piacenza, Lodi and Cremona are the neighboring cities of Codogno, the first outbreak, which, moreover, had closed the hospital almost immediately, so patients from Codogno and neighboring areas turned to the hospitals of Piacenza, Lodi and Cremona. In addition, in the early days, deaths were recorded based on where the anti-gen tests were made and not based on the patient's areas of residence. Those cities first had to deal with the emergency starting with patient 1 diagnosed on February 20, 2020. These are the cities that had to manage the surprise effect. In most other cities, hospitalizations and deaths started at least a week later, probably giving ASL more time to organize. Moreover, Piacenza, Lodi and Cremona were also not included in the first red zone, thus suffering from the spreading effects of the epidemic but with lower immediate protection measures.

\subsection{Cumulative cases and cumulative rates by province}

\begin{figure}[ht!]
\centering
    \begin{subfigure}{1\textwidth}
        \centering
        \includegraphics[width=\linewidth]{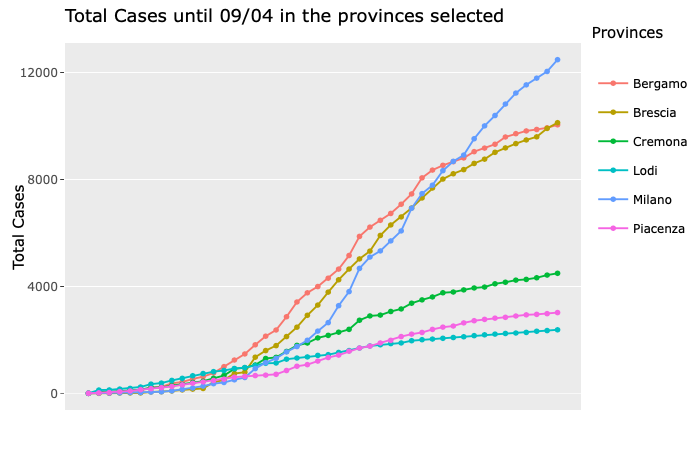}
        \caption{}\label{fig:fig_a1}
    \end{subfigure} %
    \begin{subfigure}{1\textwidth}
        \centering
        \includegraphics[width=\linewidth]{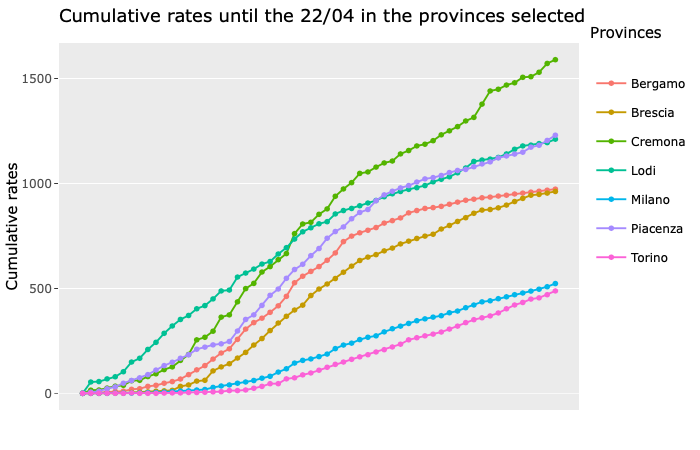}
        \caption{}\label{fig:fig_b1}
    \end{subfigure} %
\caption{Provinces most affected by Covid-19 in Italy. (\subref{fig:fig_a1}): total cases for Milan, Bergamo, Brescia, Cremona, Lodi and Piacenza provinces, (\subref{fig:fig_b1}): cumulative rates for the same provinces and Torino province}
\label{plot1}
\end{figure}

\noindent From the 'Time series' menu, it is possible to select multiple provinces and plot the series of their cumulative cases per 100,000 inhabitants (Figure \ref{plot1}(a)). It is also possible to plot the series of the cumulative rates of multiple selected provinces (Figure \ref{plot1}(b)).

\noindent Provinces to be compared with one of the two measures can also belong to different regions. In the example in Figure \ref{plot1}(a) the time series of the provinces of Milan, Bergamo, Brescia, Cremona, Lodi and Piacenza in the case of total cases are represented, whereas in Figure\ref{plot1}(b) the time series of the same provinces together with the province of Torino in the case of cumulative. These are some of the provinces most affected by Covid-19 in Italy. 

\noindent It can be seen that, considering the total cases, among the provinces represented in Figure \ref{plot1}(a), the most affected province is Milan  with 17,000 cases, followed by Brescia, Turin and Bergamo and then by Cremona, Piacenza and Lodi. On the other hand, considering the cumulative rates  as in Figure \ref{plot1}(b), Cremona stands on top with 1,590 cases per 100,000 inhabitants, followed by Piacenza and Lodi with 1,230 and 1,211, and then by Bergamo, Brescia and finally by Torino with 470 and Milan with 523 cases per 100,000 inhabitants.

\subsection{Cumulative cases and cumulative rates by region}
\begin{figure}[ht!]
\centering
    \begin{subfigure}{1\textwidth}
        \centering
        \includegraphics[width=\linewidth]{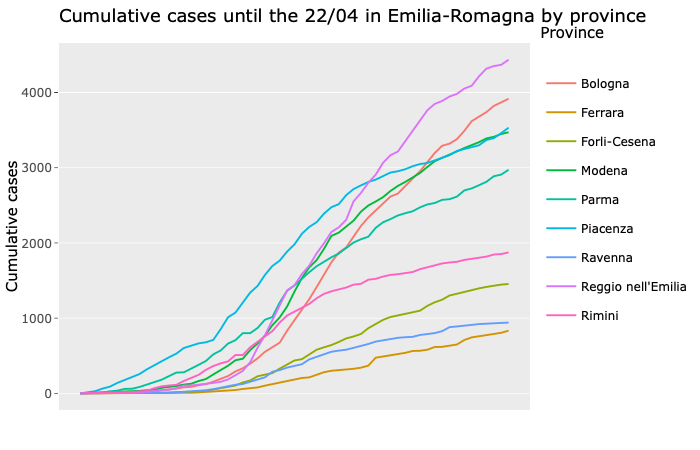}
        \caption{}\label{fig:fig_a}
    \end{subfigure} %
    \begin{subfigure}{1\textwidth}
        \centering
        \includegraphics[width=\linewidth]{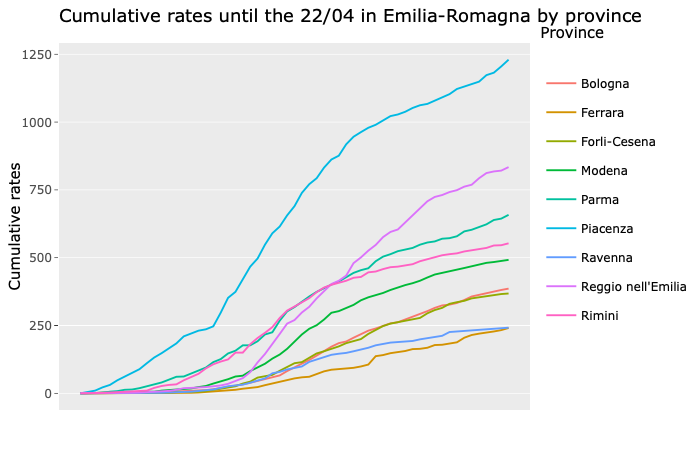}
        \caption{}\label{fig:fig_b}
    \end{subfigure} %
\caption{Emilia-Romagna provinces (\subref{fig:fig_a}): cumulative cases; (\subref{fig:fig_b}): cumulative rates}
\label{plot2}
\end{figure}

\noindent Figure \ref{plot2} shows an example of time series of provinces within the same region. The region is in this case Emilia-Romagna. The comparison is made on the basis of the cumulative cases and cumulative rates per 100,000 inhabitants. 

\noindent Notice that the trajectory of the provinces within the same region are very different. Therefore, it is difficult to think of a model that represents the whole region.
\noindent Considering the total cases in \ref{plot2}(a), Reggio Emilia, with 4,437 case, seems to be the most affected province in Emilia-Romagna. However, if one looks at the cumulative rates represented in \ref{plot2}(b), the province of Piacenza stands on top of the league table with a trajectory quite apart from those of the other provinces in the region. Piacenza is on the border between Emilia-Romagna and Lombardia and is very close to the city of Codogno in the Lodi province which was the first epicenter of the Covid-19 outbreak in Italy at the end of February 2020, so the contagion may have spread much faster there than in the rest of the Emilia-Romagna region. Newspapers reported that the so-called 'patient 1' (i.e. the first patient detected with the infection in Codogno) had multiple contacts with people from the Piacenza province since a few weeks before being hospitalized. In addition, in the last week of February many positive sick patients from the 'Basso Lodigiano' (i.e. the area of the initial 'red zone') were admitted at the main Piacenza hospital. 
 
\subsection{Comparison between weekly deaths in January-April 2015-2020 and Covid-19 deaths in March 2020}

\begin{figure}[ht!]
 \includegraphics[width=1\linewidth]{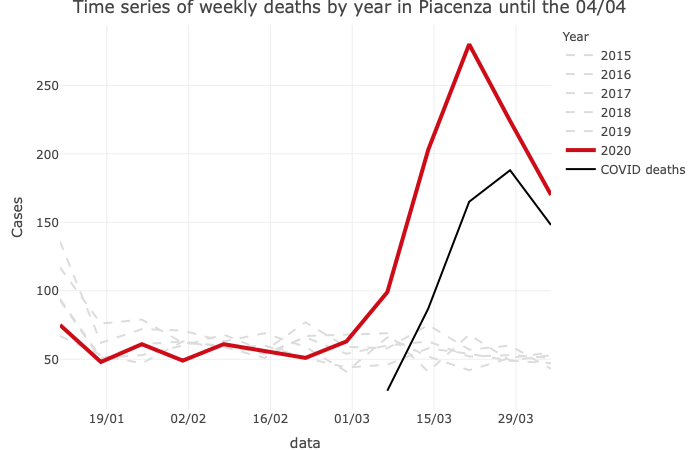}
 \caption{Weekly overall deaths in Jan-Apr, 2015-2020, and Covid-19 deaths, Mar 2020 in Piacenza province}
\label{plot3}
\end{figure}

\noindent Figure \ref{plot3} reports for a selected province the weekly series of deaths from January to April for 2015 to 2020 (red line: overall deaths in Jan-Apr 2020; grey lines: overall deaths in Jan-Apr 2015-2019; black line: Covid-19 deaths in March 2020\footnote{Source for overall deaths: ISTAT (\url{https://www.istat.it/it/archivio/240401}).}). 


\noindent Notice that \textit{data from ISTAT include only deaths from a proportion of municipalities in the province}. The number of deaths made available by ISTAT is not for a random sample of municipalities, but a selection of them (1,689 out of 5,909 municipalities) obtained according to the completeness and timeliness of the information collected, as well as the statistical criteria described in the explanatory note available by ISTAT\footnote{\url{https://www.istat.it/it/archivio/240401}.}, \textit{which should be read very carefully before using the data}. In its web page, ISTAT shows graphs at the municipal level and specifies that the data cannot simply be aggregated at the provincial, regional or national level. We show this plot for purely exploratory purposes. 

\noindent  In addition, the weekly series of deaths reported for Covid-19 is plotted in black for provinces for which we managed to obtain death data.

\noindent In Figure \ref{plot3} the Piacenza province time series of the overall deaths and Covid-19 deaths are plotted. The number of municipalities included in the overall deaths time series is 25 (out of 46).

\noindent  For the Piacenza province we have reconstructed the series of deaths for Covid-19 using the information provided in press reports taken from the site of the Emilia-Romagna region authority.  

\noindent We can firstly note that there is a significant increase in the overall deaths in 2020 (red line) starting from the first week of March 2020. The total deaths from Covid-19 declared for the whole province of Piacenza (red line) are however much less than those recorded by ISTAT which, however, only take into account a little more than a half of the municipalities. In the week from 15/03 to 21/03 280 deaths were reported in the ISTAT series and 165 deaths were reported in the Covid-19 series. In the week from 22/03 to 28/03 224 deaths were reported in the  ISTAT series and 188 deaths were reported in the Covid-19 series. In the week from 29/03 to 04/04 170 deaths were reported in the ISTAT series and 148 deaths were reported in the Covid- 19 series. The gap between the two series resulted decreasing at the end of March 2020. 

\noindent 
Two important questions cannot be fully answered here: 
\begin{itemize}
\item are the additional deaths really due to Covid-19 but these deaths wasn't declared as such because it was not possible to carry out the anti-gen tests completely for at least part of the dead people?
\item is there a generalized increase in deaths due to the fact that the medical staff and hospital wards were all busy in that period and therefore people with other serious conditions were not treated appropriately because of the Covid-19 emergency?
\end{itemize}
Probably both the above events have happened, but certainly there have been an unusual number of deaths in that period caused directly or indirectly by the Covid-19 outbreak.

\section{An adjusted SIRD model}
The SIRD model is a compartmental model used in epidemiology to design the spread of a disease that divides the population into four different groups: susceptible, currently infected, recovered, and deaths \cite{Kermack,Newman}. The size of the population is given by the sum of these four variables and it is supposed to be considered constant. The parameters governing the model are the transmission rate, the recovery rate, and the mortality rate. 
\\

Another important parameter that wholly describes the spread of an outbreak is the basic reproduction number $(R_0)$, that is computed as the ratio between the transmission rate and the sum of recovery and specific mortality rate. It is the expected number of individuals that are directly infected by one infected individual, in a population where everyone is susceptible to infection. If it is less than 1, the epidemic will eventually be controlled; if it is larger than 1, the transmission of the disease will increase in the population. 
\\

Building on Chen et al. (2020) work \cite{Chenetal}, a time-dependent model is proposed in the dashboard (section 'SIRD models') in order to let the parameters change over time. More technical details of the model can be founded in \cite{Ferrari2020}. By allowing the parameters to vary over time, the effect of containment measures, such as lock-downs, can be somewhat included in the model. Moreover, recovery and mortality rate are likely to depend on the pressure under which hospitals are in, which increases sharply at the beginning of an epidemic (i.e. when a high mortality rate is reported) and then relaxes after the health system capacity is enhanced. The number of recovered patients is obtained proportionally from that of the region where the province is in. The reason for this is that patient treatment for the illness due to Covid-19 could be considered more uniform across the various provinces (with almost the same recovery rate across provinces within the region) than for the number of deaths. In any case we are working on obtaining the series of provincial recovered cases in the same way we obtained the series of the Covid-19 deaths, i.e. from press conferences and official bulletins of regional authorities.
\\

\begin{figure}[ht!]
\centering
\includegraphics[scale=0.6]{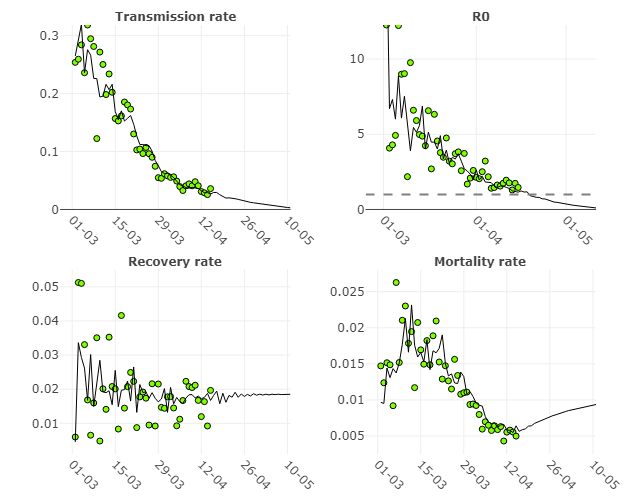}
\caption{Parameters' real values and model predictions - Italy}
\label{fig:sird_italy1}
\end{figure}

Figure \ref{fig:sird_italy1} shows the development of these parameters at the national level from the beginning of the outbreak: points indicate real data values while the solid black lines show the model predictions. The dashed line in the $R_0$ plot represents the threshold where $R_0$ is equal to 1.
\\

Given the considerable variability of the provincial data, the model is trained on the regional-aggregated data that show smoother trends. Another reason for this choice is that, although lock-downs have been declared almost simultaneously in every Italian province, the virus has spread irregularly in different geographical zones, with the southern areas reporting considerably fewer cases.
\\

\begin{figure}[ht!]
\centering
\includegraphics[scale=0.4]{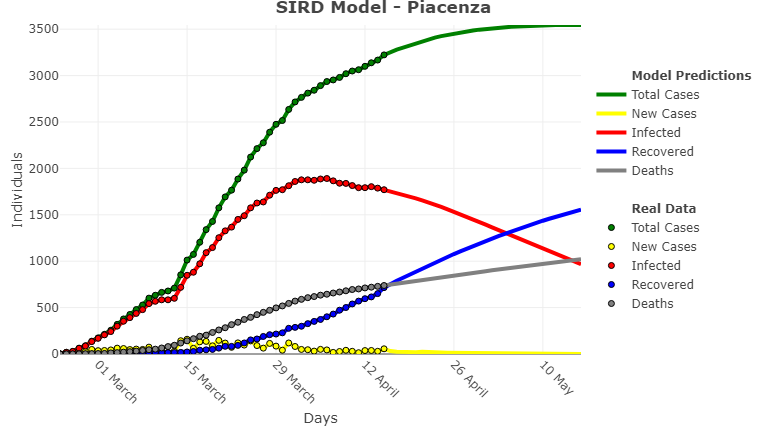}
\caption{SIRD model applied to Piacenza province (number of lags: 7)}
\label{fig:sird_italy2}
\end{figure}

In Figure \ref{fig:sird_italy2}, an example on the Piacenza province is shown. Piacenza is one of the provinces in northern Italy with the highest cumulative rate index (i.e. the ratio between reported cases and total population).
\\

\begin{figure}[ht!]
\centering
\includegraphics[scale=0.4]{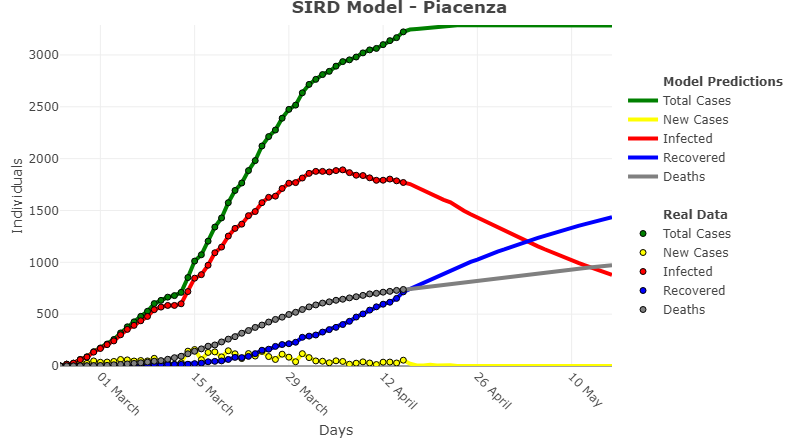}
\caption{SIRD model applied to Piacenza province (number of lags: 10)}
\label{fig:sird_italy3}
\end{figure}

The number of lag days represents the number of previous values included in the auto-regressive models used to estimate each parameter's evolution. This parameter is set by default to 7 but in the dashboard it can be changed within a range: as the number of employed lags  grows, the predictions show a more optimistic scenario. Figure \ref{fig:sird_italy3} shows the model predictions using a number of lags equal to 10, applied to the Piacenza province.
\\

Note that this is a preliminary model whose accuracy tends to decrease as predictions are done further ahead in the future. Adjustments and improvements are currently being studied and results will be updated regularly in the dashboard. \\

\section{Discussion and future work}
Official data from the Italian Ministry of Health on the Covid-19 outbreak in Italy presents many issues mainly related to delay in reporting new cases and deaths, incongruities (negative values in the series), and missing data. For example the number of deaths at province level is not reported in the official data repositories. Menchetti and Noirjean \cite{Menc} reported widely on the flaws and biases of these official data. The problem of unreliable data becomes even more cogent with epidemiological models, both deterministic and stochastic, when many parameters should be estimated on the basis of these unreliable data, especially for long-range estimates which are even more important for an outbreak with such dramatic consequences the whole world is experiencing. In this context, future work will be dedicated to obtaining the series of provincial recovered cases in the same fashion we obtained the series of provincial deceased cases, i.e. scraping websites of regional authorities.

It is our personal opinion that caution should always be used when modeling and predicting the outcomes of events like this outbreak which has many different facets to be considered and changes continuously during the time, see also \cite{Bartoszek}. 

However, the difficulty in obtaining acceptable estimates should not be an "excuse" not to try to contribute to the knowledge of phenomenons like this one which is changing all our lives. Therefore our efforts to obtain more reliable estimates will continue by exploring improved adjusted models, included more SIRD-type models (like the SIERD model, an extension of the SIR model to consider exposed people) and a stochastic models taking into consideration the uncertainty in the estimates.

\end{document}